\documentclass[12pt]{cernart}
\usepackage{times}
\usepackage{graphicx}
\newcommand{\beq}  {\begin{equation}}
\newcommand{\eeq}  {\end{equation}}
\newcommand{\bmath}{\begin{eqnarray}}
\newcommand{\emath}{\end{eqnarray}}

\def\lapproxeq{\lower .7ex\hbox{$\;\stackrel{\textstyle<}{\sim}\;$}}
\def\gapproxeq{\lower .7ex\hbox{$\;\stackrel{\textstyle>}{\sim}\;$}}
\def\be{\begin{equation}}
\def\ee{\end{equation}}
\def\GeV{\mbox{GeV}}
\begin{document}
\begin{titlepage}
\docnum{CERN--PH--EP/2005--001}
\addtocounter{footnote}{1}
\renewcommand{\thefootnote}{\fnsymbol{footnote}}
\hbox to \hsize{\hskip123mm\hbox{18 January 2005}\hss}
\hbox to \hsize{\hskip123mm\hbox{revised\footnotemark\ 22 March 2005}\hss}
\footnotetext{author list updated, details added, Fig.\ 4 modified}

\vspace{1cm}
\title{\Large Measurement of the Spin Structure of the Deuteron
         in the  DIS Region}

%
\author{\large The COMPASS Collaboration}

\vspace{3cm}
\begin{abstract}
We present a new measurement of the longitudinal spin asymmetry $A_1^d$
and the spin-dependent structure function  $g_1^d$ of the deuteron  in the
range $1~\GeV^2 < Q^2 < 100~\GeV^2$ and $0.004<  x  <0.7$.
The data were obtained by the COMPASS experiment at CERN
using a 160~GeV polarised muon beam and a large polarised $^6$LiD target.
The results are
in  agreement with those from previous experiments and improve considerably the
statistical accuracy in the region $0.004 <  x  < 0.03$.
\vfill
\submitted{(Submitted to Physics Letters B)}
\end{abstract}

\newpage
%
%
\begin{Authlist}
{\large  The COMPASS Collaboration}\\[\baselineskip]
%
%
E.S.~Ageev\Iref{protvino},
V.Yu.~Alexakhin\Iref{dubna},
Yu.~Alexandrov\Iref{moscowlpi},
G.D.~Alexeev\Iref{dubna},
A.~Amoroso\Iref{turin},
B.~Bade\l ek\Iref{warsaw},
F.~Balestra\Iref{turin},
J.~Ball\Iref{saclay},
G.~Baum\Iref{bielefeld},
Y.~Bedfer\Iref{saclay}
% ...
\end{Authlist}
%
%
\Instfoot{bielefeld}{ Universit\"at Bielefeld, Fakult\"at f\"ur Physik, 33501 Bielefeld, Germany\Aref{d}}
\Instfoot{bochum}{ Universit\"at Bochum, Institut f\"ur Experimentalphysik, 44780 Bochum, Germany\Aref{d}}
\Instfoot{bonniskp}{ Universit\"at Bonn, Helmholtz-Institut f\"ur  Strahlen- und Kernphysik, 53115 Bonn, Germany\Aref{d}}
\Instfoot{bonnpi}{ Universit\"at Bonn, Physikalisches Institut, 53115 Bonn, Germany\Aref{d}}
\Instfoot{brno}{Institute of Scientific Instruments, AS CR, 61264 Brno, Czech Republic\Aref{e}}
\Instfoot{burdwan}{ Burdwan University, Burdwan 713104, India\Aref{g}}
\Instfoot{calcutta}{ Matrivani Institute of Experimental Research \& Education, Calcutta-700 030, India\Aref{h}}
\Instfoot{dubna}{ Joint Institute for Nuclear Research, 141980 Dubna, Moscow region, Russia}
\Instfoot{erlangen}{ Universit\"at Erlangen--N\"urnberg, Physikalisches Institut, 91054 Erlangen, Germany\Aref{d}}
\Instfoot{freiburg}{ Universit\"at Freiburg, Physikalisches Institut, 79104 Freiburg, Germany\Aref{d}}
\Instfoot{cern}{ CERN, 1211 Geneva 23, Switzerland}
\Instfoot{heidelberg}{ Universit\"at Heidelberg, Physikalisches Institut,  69120 Heidelberg, Germany\Aref{d}}
\Instfoot{helsinki}{ Helsinki University of Technology, Low Temperature Laboratory, 02015 HUT, Finland  and University of Helsinki, Helsinki Institute of  Physics, 00014 Helsinki, Finland}
\Instfoot{licerec}{Technical University in Liberec, 46117 Liberec, Czech Republic\Aref{e}}
\Instfoot{lisbon}{ LIP, 1000-149 Lisbon, Portugal\Aref{f}}
\Instfoot{mainz}{ Universit\"at Mainz, Institut f\"ur Kernphysik, 55099 Mainz, Germany\Aref{d}}
\Instfoot{miyazaki}{University of Miyazaki, Miyazaki 889-2192, Japan\Aref{i}}
\Instfoot{moscowlpi}{Lebedev Physical Institute, 119991 Moscow, Russia}
\Instfoot{munichlmu}{Ludwig-Maximilians-Universit\"at M\"unchen, Department f\"ur Physik, 80799 Munich, Germany\Aref{d}}
\Instfoot{munichtu}{Technische Universit\"at M\"unchen, Physik Department, 85748 Garching, Germany\Aref{d}}
\Instfoot{nagoya}{Nagoya University, 464 Nagoya, Japan\Aref{i}}
\Instfoot{praguecu}{Charles University, Faculty of Mathematics and Physics, 18000 Prague, Czech Republic\Aref{e}}
\Instfoot{praguectu}{Czech Technical University in Prague, 16636 Prague, Czech Republic\Aref{e}}
\Instfoot{protvino}{ State Research Center of the Russian Federation, Institute for High Energy Physics, 142281 Protvino, Russia}
\Instfoot{saclay}{ CEA DAPNIA/SPhN Saclay, 91191 Gif-sur-Yvette, France}
\Instfoot{telaviv}{ Tel Aviv University, School of Physics and Astronomy, 
              69978 Tel Aviv, Israel\Aref{j}}
\Instfoot{triestictp}{ ICTP--INFN MLab Laboratory, 34014 Trieste, Italy}
\Instfoot{triest}{ INFN Trieste and University of Trieste, Department of Physics, 34127 Trieste, Italy}
\Instfoot{turin}{ INFN Turin and University of Turin, Physics Department, 10125 Turin, Italy}
\Instfoot{warsaw}{ So{\l}tan Institute for Nuclear Studies and Warsaw University, 00-681 Warsaw, Poland\Aref{k} }
\Instfoot{warsawtu}{ Warsaw University of Technology, Institute of Radioelectronics, 00-665 Warsaw, Poland\Aref{l} }
\Anotfoot{a}{Also at IST, Universidade T\'ecnica de Lisboa, Lisbon, Portugal}
\Anotfoot{b}{Also at University of East Piedmont, 15100 Alessandria, Italy}
\Anotfoot{c}{On leave of absence from JINR Dubna}               
\Anotfoot{d}{Supported by the German Bundesministerium f\"ur Bildung und Forschung}
\Anotfoot{e}{Suppported by Czech Republic MEYS grants ME492 and LA242}
\Anotfoot{f}{Supported by the Portuguese FCT - Funda\c{c}\~ao para
               a Ci\^encia e Tecnologia grants POCTI/FNU/49501/2002 and POCTI/FNU/50192/2003}
\Anotfoot{g}{Supported by UGC-DSA II grants, Govt. of India}
\Anotfoot{h}{Supported by  the Shailabala Biswas Education Trust}
\Anotfoot{i}{Supported by the Ministry of Education, Culture, Sports,
               Science and Technology, Japan}
\Anotfoot{j}{Supported by the Israel Science Foundation, founded by the Israel Academy of Sciences and Humanities}
\Anotfoot{k}{Supported by KBN grant nr 621/E-78/SPUB-M/CERN/P-03/DZ 298 2000 and
               nr 621/E-78/SPB/CERN/P-03/DWM 576/2003-2006}
\Anotfoot{l}{Supported by  KBN grant nr 134/E-365/SPUB-M/CERN/P-03/DZ299/2000}

\vfill

\hbox to 0pt {~}
%
%
\end{titlepage}
%
%
%
\noindent
Since the surprising result obtained for the spin structure function of the proton by the EMC \cite{EMC}, the determination 
of the longitudinal spin structure of the proton  and the neutron has   remained one of 
the important issues in  particle  
physics \cite{averett}. The spin structure functions are used to test the Bjorken sum rule
and  
to determine quark and gluon polarisations from the QCD evolution equations \cite{aac}. They are also used as constraints
in the derivation of the polarisation of quarks of different flavour from semi-inclusive
asymmetries \cite{semiinc2,semiinc4}. 

Here we report on the first results from the COMPASS experiment at CERN on the deuteron spin asymmetry $A_1^d$ 
and the spin-dependent structure function $g_1^d$  in 
the deep inelastic scattering (DIS) region,  
covering the range 1~GeV$^2$ to 100~GeV$^2$ in  the photon virtuality $Q^2$ and 0.004 to 0.7
in the Bjorken scaling variable $x$. 

The COMPASS spectrometer    
is located in the same muon beam line as the former SMC experiment  and  covers a similar kinematic region for inclusive 
reactions. However, it uses a higher intensity muon beam of 160~GeV, a longitudinally or transversely
polarised target made of $^6$LiD, and a new two-stage 
spectrometer. A general description of the experiment has been presented in Ref.~\cite{gkm} and only the most relevant
elements for the present analysis will be mentioned below.
 The data in the longitudinal configuration 
 taken  in 2002 and  2003 correspond to luminosities of 
about 600~pb$^{-1}$ and 900~pb$^{-1}$,
respectively. 

The experiment was performed at the M2 muon beam line of the CERN SPS.
The muons originate from the decay of $\pi$ and K mesons produced by the 400~GeV proton beam on a
primary beryllium target.
The  $\mu^+$ intensity is $2\cdot10^8$ per 
spill of 4.8~s with a cycle time of $16.8$~s.  
The beam profile presents a Gaussian core and a large non-Gaussian tail due to halo muons.
The beam has a nominal energy of 160~GeV and is focused at the target centre, with a  spread of 7~mm (r.m.s.) 
and a momentum spread of $\sigma_p/p=0.05$ for the Gaussian core. 
The momentum of each muon is measured upstream of the experimental area in a beam momentum station  consisting of
five (four in the year 2002)  planes of scintillator strips  with 
a dipole magnet in between. 
The precision of the momentum determination is typically $\Delta p/p=0.003$. 
The incoming muon direction and position are measured 
by small scintillating fibre hodoscopes  
and  silicon microstrip 
detectors \cite{SciFi,Si}.
The space resolution is about 0.12~mm for the fibres and 0.015~mm for the microstrips, and the direction of the 
incoming muon  is measured 
with a precision of 30~${\rm\mu}$rad.

 The  polarisation $P_B$ of the beam muons was determined 
by a Monte Carlo program  modelling in detail the phase space of the parent hadrons and decay muons, as well as their 
propagation through the beam transport system \cite{mupolMC}. 
Within a precision of about 0.04 the calculated values are consistent with the 
 polarisation measurements performed by the SMC at 
100 and 190~GeV
\cite{mupolSMC}.
For the present experiment the model gives
a polarisation of the  muon  varying with its  energy   from
  $-0.57$ at 140~GeV to $-0.86$ at 180~GeV with a mean value of
  $-0.76$.                                                    

\begin{figure}[here]
  \begin{center}
    \includegraphics[width=1.0\textwidth,clip]{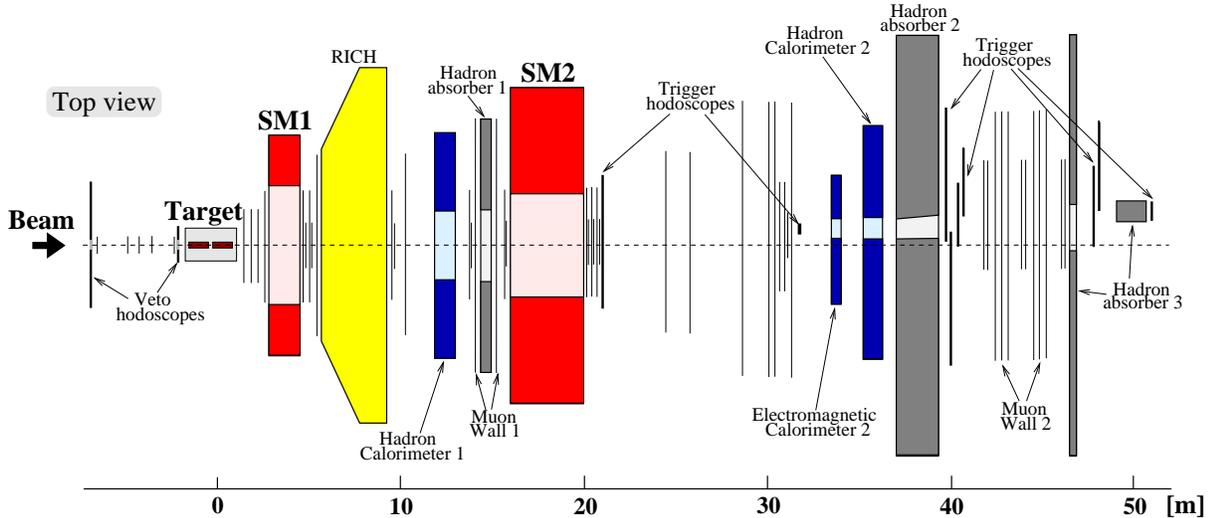}
    \caption{Layout of the COMPASS spectrometer  used in 2003. The configuration was identical in
     2002 except for Electromagnetic Calorimeter 2 which was not included in the read-out.
     The thin vertical lines represent the tracking detectors.
      }
\label{fig:layout}
  \end{center}
\end{figure}
The target is located inside the solenoid magnet previously used by the SMC experiment \cite{smc_ta},
which provides a field of 2.5~T along the beam direction. The magnet aperture seen from the upstream
end of the target is $\pm 70$~mrad.
The  target consists of two cells, each 60~cm 
long and 3~cm in diameter, separated by 10~cm. 
They are filled with $^6$LiD which is used as  deuteron target material and longitudinally polarised with
dynamic nuclear polarisation (DNP) \cite{target}. 
The two  cells are  polarised in opposite directions so that 
data from both spin directions are recorded at the same time.   
The polarisation is measured by NMR coils with a relative precision of about 5\%
\cite{compassNMR}.
The typical polarisation values obtained after
a build-up time of about 5 days are $+0.53$ and $-0.50$. 
The spin directions in the two target cells                                
are reversed every 8 hours by rotating the magnetic field direction.              
In this way, fluxes and acceptances cancel out in the calculation of spin asymmetries,
provided that the ratio of acceptances remains unchanged  after spin reversal.  
In order to minimise possible acceptance effects related to the orientation of the solenoid field,
the sign of the polarisation in each target cell  is also reversed several times per year by changing the DNP microwave frequencies.

The COMPASS spectrometer
(Fig.~\ref{fig:layout}) 
is  designed to reconstruct the scattered muons and the 
produced hadrons in wide momentum and angular ranges. It is divided in two stages associated with two dipole magnets, 
SM1 and SM2. The first one is a large-aperture magnet, with 
a field integral of 1~Tm along the beam line, which accepts charged particles of 
momenta larger than 0.4~GeV. The second magnet, SM2, has  
a field integral of 4.4~Tm and accepts particles of momenta 
larger than 4~GeV. 
Different types of tracking detectors are used  to cope
with the rapid increase of the particle rate  from the outside to 
the central beam region. The beam region downstream of the target is covered by scintillating fibre detectors \cite{SciFi},  
the region near to the beam  by micromesh gaseous chambers  \cite{micromegas} and gas electron multiplier chambers \cite{gem}. 
The intermediate region, further away from the beam line, is covered by drift chambers  and multiwire proportional chambers.
Large-angle tracking is mainly provided by straw detectors  \cite{straw} and by 
large drift  chambers.  
The identification  of muons is based on the fact that they are observed behind hadron absorbers. 
Two `muon wall' detectors are used: the first one, 
located in front of SM2, 
consists of two stations of  Iarocci-type chambers with 
an iron layer in between and detects muons outside the aperture of SM2;
the second one, installed at the end of the spectrometer,
is composed of drift tubes and detects the muons which passed through SM2.
Hadrons are detected by
 two large  iron-scintillator sampling calorimeters, installed in front of the absorbers
and shielded to avoid electromagnetic contamination.

\begin{figure}[tb]
  \begin{center}
    \includegraphics[width=0.49\textwidth,clip]{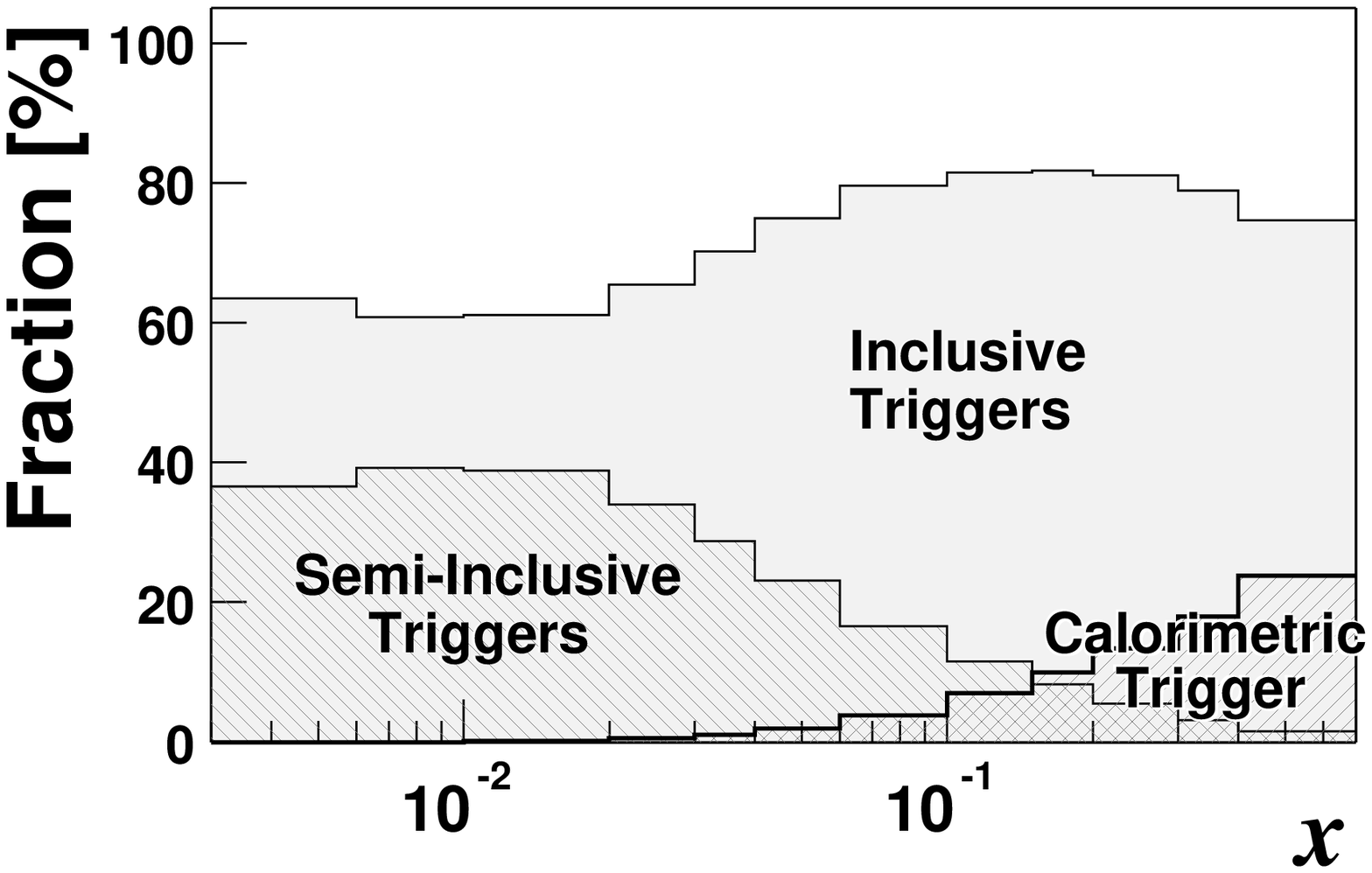}
    \includegraphics[width=0.49\textwidth,clip]{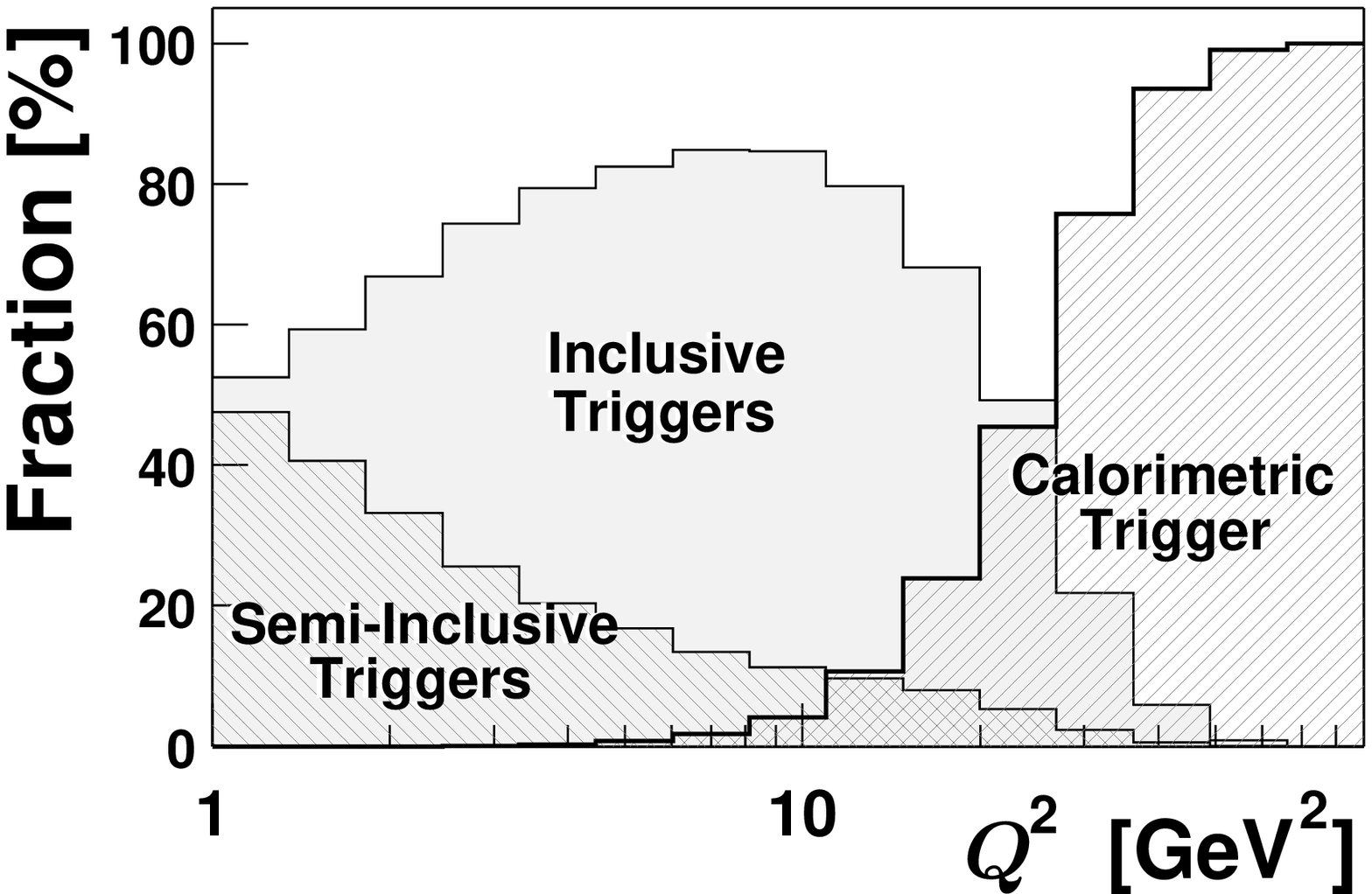}
    \caption{Fraction of inclusive, semi-inclusive, and calorimetric triggers
      in the final data sample (2002 and 2003) as a function of $x$ (left) and $Q^2$ (right). Events are counted with the
      weight they carry in the asymmetry calculation [Eq.~(7)].}                 
\label{fig:xq2}
  \end{center}
\end{figure}
The data recording system is activated by a combination of signals indicating the presence of a
scattered muon at a given angle or in a given energy range. In most  DIS events ($Q^2 > 1~\GeV^2$),
the scattered muon is identified by coincidence signals in the trigger hodoscopes, 
that define
its direction behind SM2. Several veto counters installed upstream of the target
are used to avoid triggers due to halo muons.
In addition to this inclusive trigger mode, which was commonly used in the previous CERN muon experiments,
several semi-inclusive triggers select  
events fulfilling requirements based on the
muon energy loss and on the presence of a hadron  signal in the calorimeters \cite{trigger}.
Calorimeter signals due to
halo muons are rejected by requiring the presence of at least one cluster with an energy deposit exceeding
three times the average value expected for a muon. This condition provides a trigger efficiency of
more than 80\% for events with total hadronic energy $E_{had} >  30~\GeV$.
In a part of the 2003 data taking, the acceptance was further extended towards  high $Q^2$ values by the
addition of a standalone calorimetric trigger in which no condition is set for the scattered muon but an 
energy deposit in the hadron calorimeter exceeding 9 times the typical muon response is required.
The semi-inclusive and calorimetric triggers thus select a sample of hadronic events which are    
analysed in parallel with the inclusive sample.
The relative contributions of  the different trigger types are shown as a function of $x$ and  
 $Q^2$  in Fig.~\ref{fig:xq2}.
The fraction of inclusive triggers, where the selection criteria refer only to the scattered muon, varies from  60\% to
75\% over the  range of $x$
(events satisfying simultaneously inclusive and non-inclusive trigger conditions
are counted as inclusive). The semi-inclusive triggers account for
about 40\% of the data at low $x$ and decrease steadily for $x > 0.02$, while the contribution of
the standalone calorimetric trigger starts around $x = 0.02$ and reaches 30\% in the highest $x$ bin.

Larger variations of the different contributions are observed as a function of $Q^2$:  
the inclusive triggers account for 80\% of the events at medium $Q^2$ (3--15~GeV$^2$), while             
the standalone
calorimetric trigger
becomes  dominant for  $Q^2 > 30~\GeV^2$.  

In order to eliminate spurious triggers as well as badly or partially reconstructed events,
a reconstructed
interaction point connected to a beam muon and to a scattered muon is required for all events.
In addition, the presence of a hadron track at the interaction point is required for the semi-inclusive and
standalone calorimetric triggers.
The track reconstruction efficiency was found to
be about 95\% for  scattered muons and for high-energy hadrons ($E > 30~\GeV$) 
that were generated in a Monte Carlo simulation, tracked through
the spectrometer, and analysed in the same way as the data.
The direction of tracks reconstructed at the interaction point is determined with a precision
better than 0.2~mrad and the momentum resolution for scattered muons is about 0.5\%.

As the COMPASS trigger setup is predominantly intended for the study of quasi-real
photon interactions, DIS events represent only a small fraction of the data sample.
 The combination of  cuts on the photon virtuality
 ($Q^2>1~\GeV^2$),  the fraction of energy carried away by the virtual photon
($0.1<y<0.9$), and  the requirement that the interaction take place within one of the target cells
results in a reduction factor of about 20.
In addition, the  incoming muon momentum is required to be in the interval $140~\GeV<p_{\mu}<180~\GeV$  
and, in order to equalise fluxes seen by the two target cells,  
its trajectory is required to cross entirely both target cells.                                    
For consistency, in events triggered by hodoscope signals,
 it is also verified that the reconstructed scattered muon hits the hodoscopes
that have generated the event trigger. The resulting sample amounts to about $34 \cdot 10^6$ events
with a fraction of 71\% of the data collected in 2003. 

\begin{figure}[t]
  \begin{center}
    \includegraphics[width=0.6\textwidth,clip]{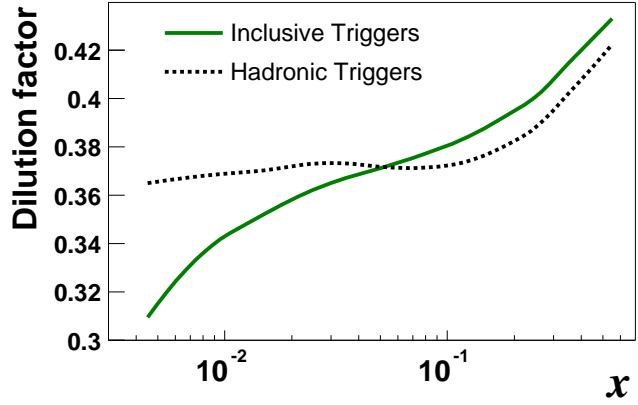}
    \caption{The dilution factor $f$ of the $^6$LiD polarised target as a function of $x$ for inclusive
     and  hadronic events. The dilution due to radiative effects on the deuteron is included.
     The values quoted for each $x$ bin are averaged over the kinematic range of the corresponding triggers.} 
\label{fig:dilut}
  \end{center}
\end{figure}
The cross-section asymmetry $A^d = (\sigma^{\uparrow \downarrow} - \sigma^{\uparrow \uparrow}) /
(\sigma^{\uparrow \downarrow} + \sigma^{\uparrow \uparrow})$, for antiparallel 
($\uparrow \downarrow$) and parallel ($\uparrow \uparrow$)  spins of the incoming muon and the target deuteron, is
related to the virtual-photon deuteron asymmetries $A_1^d$ and $A_2^d$ by
\begin{equation}
 A^d = D (A_1^d + \eta A_2^d)~,
\end{equation}
where the factors $\eta$ and $D$ depend on the event kinematics.    
The virtual-photon depolarisation factor 
\begin{equation}
  D \simeq   \frac{y (2 - y)}{y^2 + 2 (1+R) (1-y)}
\end{equation}
depends in addition on the unpolarised structure function $R = \sigma_L/\sigma_T$. 
The longitudinal virtual-photon deuteron asymmetry is defined as
\begin{equation}
A_1^d = (\sigma_0^T - \sigma_2^T) / (2 \sigma^T)~,
\end{equation}
where $\sigma_J^T$ is the virtual-photon--deuteron absorption cross-section for total spin projection $J$ in the photon direction,
and $\sigma^T = (1/3)~(\sigma_0^T + \sigma_1^T + \sigma_2^T)$ is the total transverse photo-absorption
cross-section. 
The transverse asymmetry $A_2^d$ has been accurately measured \cite{e155_a2} and was
found to be  small. Since the kinematic factor $\eta = \frac{2(1-y)} {y (2-y)} \sqrt{Q^2}/ E_{\mu}$ is also
small in the COMPASS kinematic range, 
the second term in Eq.~(1) can be neglected, so that
\begin{equation}
A^d_1 \simeq  A^d/D~,
\end{equation}
and the longitudinal spin structure function is given by
\begin{equation}
g_1^d = \frac{F_2^d}{2~ x~(1 + R)} A_1^d~,
\end{equation}  
where $F_2^d$ is the deuteron spin-independent structure function. 
The number of events $N_i$ collected from a given target cell in a given time interval is related to the spin-independent
cross-section ${\overline \sigma}$ and to the asymmetry $A_1^d$ by
\begin{equation}
N_i = a_i \phi_i n_i {\overline \sigma} (1 + P_B P_T f D A_1^d)~,
\end{equation}
where $P_B$ and $P_T$ are the beam and target polarisations,
$\phi_i$ the incoming muon flux, 
$a_i$ the acceptance for
the  target cell, $n_i$ the corresponding number of target nucleons,
and $f$ the target dilution factor.
For a $^6$LiD target the dilution is naively expected to be of the order of 50\% because $^6$Li can be described
as an $^4$He core and a deuteron \cite{rondon}.
The dilution factor $f$ is given by the ratio of the absorption cross-sections on the deuteron to that of
all  nuclei entering  the target cells. It includes a correction for the relative polarisation of
deuterons bound in $^6$Li with respect to free deuterons.
 It also includes the dilution due to
radiative events on the deuteron, which is   taken into account by the ratio of the one-photon exchange
cross-section to the total cross-section $\rho = \sigma_d^{1 \gamma} / \sigma_d^{tot}$ 
\cite{terad}.
The values of $f$  are shown in Fig.~\ref{fig:dilut} as a function of $x$
for inclusive and hadronic events. The large difference observed at low $x$ results from the factor $\rho$
which is much smaller in the inclusive case because radiative effects in 
elastic scattering largely contribute 
in the denominator.
The dilution factors also differ slightly at high $x$ because the inclusive and standalone calorimetric
triggers cover different ranges of $Q^2$ as shown in Fig.~\ref{fig:xq2}. 

\begin{figure}[t]
  \begin{center}
    \includegraphics[width=1.0\textwidth,clip]{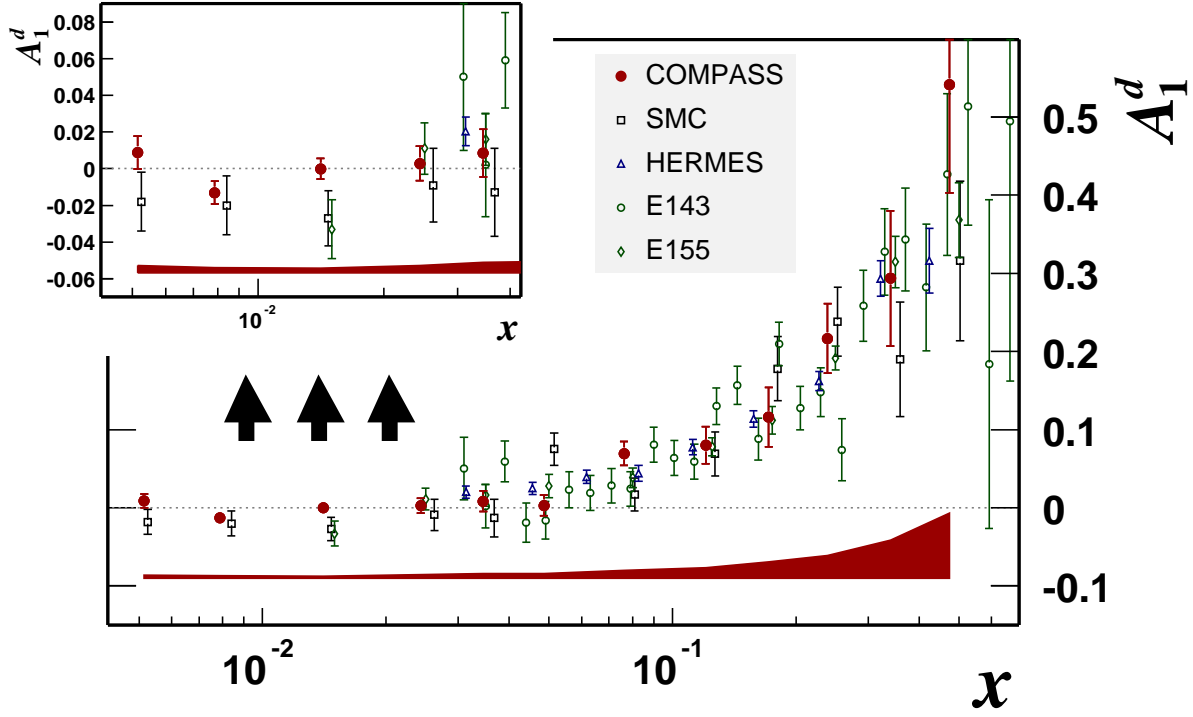}
 \caption{The asymmetry $A_1^d(x)$ as  measured in COMPASS and  previous results from SMC
     \cite{smc}, HERMES \cite{semiinc4},  SLAC E143 \cite{e143} and E155 \cite{e155}
 at  $Q^2>1~\GeV^2$. The SLAC values of $g_1/F_1$ have been
converted to $A_1$ and the E155 data corresponding to the same $x$ have been averaged over $Q^2$.
      Only statistical errors are shown with the data points.
      The shaded areas  show the size of the COMPASS systematic errors.}
\label{fig:asym}
  \end{center}
\end{figure}
The asymmetry is extracted from  data sets taken before and
after a reversal of the target spin directions. The four relations of Eq.~(6), corresponding to the two cells ($u$ and $d$) and
the two spin orientations (1 and 2), lead to  a second-order
equation in $A_1^d$ for the ratio $(N_{u,1} N_{d,2})/(N_{d,1} N_{u,2})$.
Fluxes and acceptances cancel out in this equation if 
the ratio of acceptances for the two cells
is the same before and after the reversal \cite{long_pp}.  
 In order to minimise the statistical error, all quantities used in the asymmetry calculation are evaluated
event by event with the weight factor
\begin{equation}
w = P_B f D.
\end{equation}
The  polarisation of the beam muon, $P_B$, is obtained from a parametrisation as a function of the beam  momentum.
The factors $f$ and $D$ are calculated  from the kinematic variables with the
value of $R$ taken from the NMC \cite{nmc_r} or the SLAC parametrisation \cite{slac_r} for $x$ 
below or above 0.12, respectively.
The target polarisation is not  included in the event weight  [Eq.~(7)] because it may vary in time
and generate false asymmetries. An 
average $P_T$ ($\approx 0.5$) is used for each target cell and each spin orientation.

Inclusive and hadronic events are analysed separately with the corresponding value of the dilution factor.
The additive
radiative correction to the asymmetry \cite{long_pp} has also been calculated separately
\cite{polrad} using an input parametrisation of $A_1^d$ 
fitted to
the present data. The  values obtained for inclusive and hadronic events differ  by 0.0003 in the lowest $x$-intervals and become nearly
equal at higher $x$.
  These additive corrections are negligible at low $x$
and reach a maximum value of 0.008 at high $x$. \\
The asymmetries obtained for hadronic events are statistically compatible with the inclusive ones and
their
differences do not show any hint of a systematic dependence on $x$. This observation agrees with the
Monte Carlo study of Ref.~\cite{smc} which also shows that the selection of hadronic events
has no sizeable effect on the evaluation of $A_1$ for interactions on a deuteron target
within the kinematic range and the hadron acceptance of the present experiment.

\begin{table}  [t]
\begin{center}
\caption{
Values of $A_1^d$ and $g_1^d$ with their statistical and systematical errors 
as a function of $x$ with the corresponding
average values of $Q^2$ and $y$.}

\medskip
\small
\begin{tabular}{|c|c|c|c|c|c|}
\hline
$x$ range & $\langle x \rangle$ & $\langle Q^2 \rangle$ & $ \langle y \rangle$ & $A_1^d$ & $g_1^d$ \\
 &  & $ ({\rm GeV^2}) $ &  &  &  \\
\hline
0.004$-$0.006 & 0.0051 &  1.18 & 0.76 &~$~0.009\pm0.009\pm0.004$ &~$~0.190\pm0.195\pm0.090$ \\
0.006$-$0.010 & 0.0079 &  1.53 & 0.64 & $-0.013\pm0.006\pm0.003$ & $-0.203\pm0.096\pm0.047$ \\
0.010$-$0.020 & 0.0141 &  2.28 & 0.54 & $~0.000\pm0.006\pm0.003$ & $-0.001\pm0.056\pm0.025$ \\
0.020$-$0.030 & 0.0243 &  3.38 & 0.46 &~$~0.003\pm0.009\pm0.004$ &~$~0.018\pm0.059\pm0.027$ \\
0.030$-$0.040 & 0.0345 &  4.53 & 0.43 &~$~0.008\pm0.013\pm0.006$ &~$~0.039\pm0.060\pm0.028$ \\
0.040$-$0.060 & 0.0486 &  6.08 & 0.41 &~$~0.003\pm0.013\pm0.006$ &~$~0.010\pm0.044\pm0.020$ \\
0.060$-$0.100 & 0.0762 &  8.74 & 0.38 &~$~0.069\pm0.015\pm0.010$ &~$~0.149\pm0.033\pm0.020$ \\
0.100$-$0.150 & 0.1205 & 12.9  & 0.35 &~$~0.080\pm0.024\pm0.013$ &~$~0.103\pm0.031\pm0.017$ \\
0.150$-$0.200 & 0.1717 & 17.5  & 0.34 &~$~0.116\pm0.038\pm0.021$ &~$~0.096\pm0.031\pm0.017$ \\
0.200$-$0.300 & 0.2390 & 23.9  & 0.33 &~$~0.217\pm0.045\pm0.029$ &~$~0.110\pm0.023\pm0.014$ \\
0.300$-$0.400 & 0.3401 & 34.0  & 0.33 &~$~0.294\pm0.086\pm0.048$ &~$~0.074\pm0.022\pm0.012$ \\
0.400$-$0.700 & 0.4740 & 47.5  & 0.33 &~$~0.542\pm0.139\pm0.083$ &~$~0.050\pm0.013\pm0.007$ \\
\hline
\end{tabular}
\end{center}
\end{table}
The final values of $A_1^d$ are obtained by merging the inclusive and hadronic sets weighted according to their statistical
errors.
They are listed in Table~1 with the corresponding  statistical and systematical errors
and shown     
in Fig.~\ref{fig:asym} in comparison with those obtained
by the SMC \cite{smc}, by E143 \cite{e143} and E155 \cite{e155} at SLAC,  and by HERMES \cite{semiinc4}. Good agreement is observed over
the full range of $x$. For the four points with $x < 0.03$, 
our results reduce the statistical errors 
of previous measurements by a factor of about 2.5.  

\begin{figure}[t]
  \begin{center}
    \includegraphics[width=0.8\textwidth,clip]{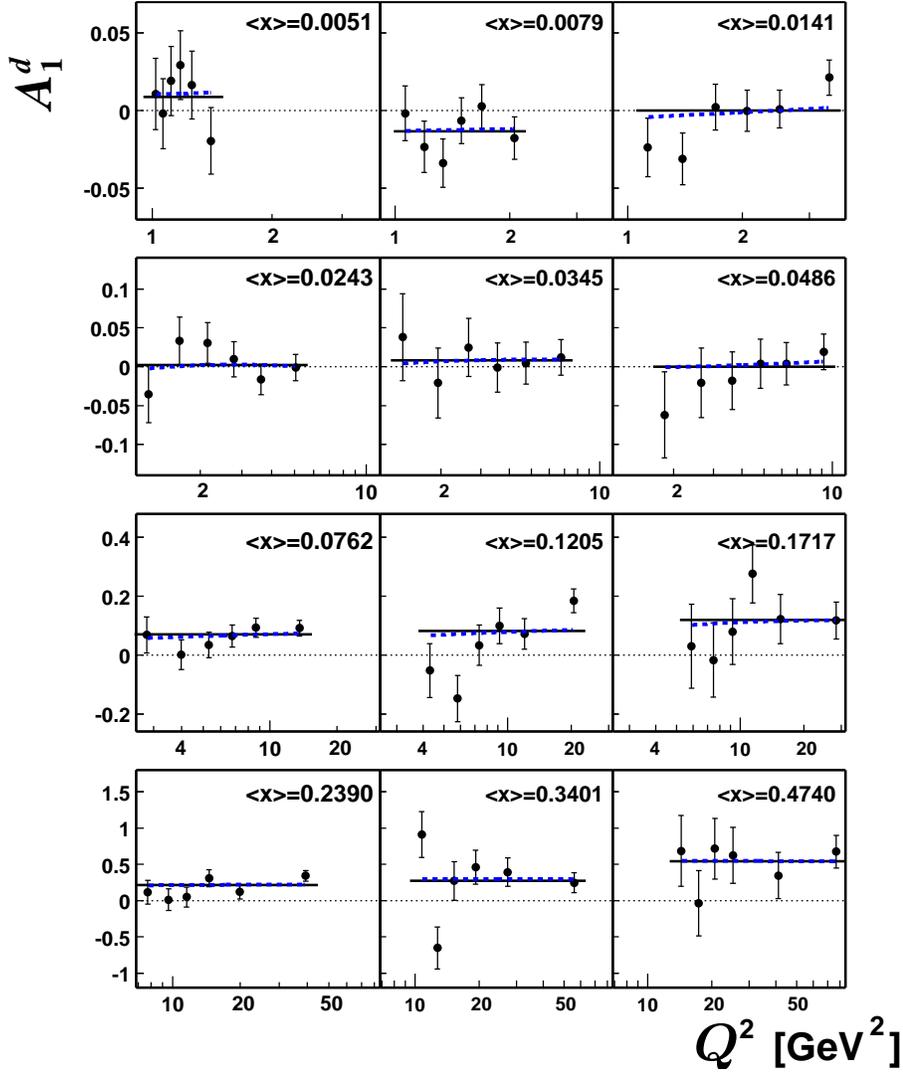}
    \caption{Values of $A_1^d$ as a function of $Q^2$ in  intervals of $x$.
     The solid lines are the results of fits to a constant; the dashed lines show the 
     $Q^2$ dependence predicted by perturbative QCD.
      }
\label{fig:a1d2}
  \end{center}
\end{figure}

Figure~\ref{fig:a1d2} shows the values of
$A_1^d$  as a function  of $Q^2$ for each interval of $x$.
The  results of fits to a constant
in each interval of $x$ are shown by the solid lines.
They  yield an average $\chi^2$-probability of about 0.5 and
do not  indicate any  $Q^2$ dependence.
Some  dependence of $A_1^d$ on $Q^2$ is expected
from perturbative QCD because the $Q^2$ evolutions of 
spin dependent and spin independent structure functions are different. However previous experiments
\cite{smc} have shown that the two $Q^2$ evolutions largely cancel out so that the values of $A_1^d$ at fixed
$x$ become nearly independent of $Q^2$. The $Q^2$ dependence predicted by the SMC fit of Ref.~\cite{smc_qcd}  
is  shown by the dashed lines in Fig.~\ref{fig:a1d2} and describes the data  equally well. 

\begin{figure}[htb]
  \begin{center}
    \includegraphics[width=0.93\textwidth,clip]{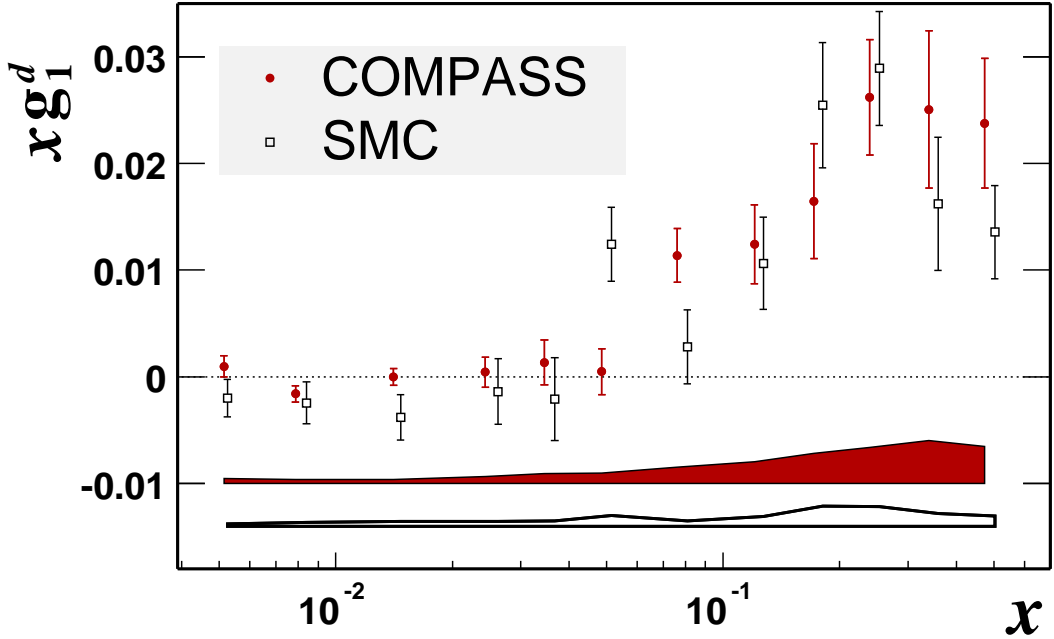}
    \caption{Values of  $x\,g_1^d(x)$ {\it vs}.\,$x$. The COMPASS points are given at the $ \langle Q^2 \rangle$  of each interval of $x$.
     The  SMC points \cite{smc} were evolved to the $Q^2$ of the corresponding COMPASS point and are slightly
     shifted to larger $x$ for clarity. Only statistical errors are shown with the data points. 
     The upper and lower bands show the COMPASS and SMC systematic errors, respectively.                     
      }
\label{fig:g1compa}
  \end{center}
\end{figure}
The systematic error on $A_1^d$ contains an overall scale uncertainty of 6.5\% due to the uncertainties
on $P_B$ and $P_T$.
The error on the dilution factor $f$,
which takes into account the uncertainty on the  target composition and
the uncertainty on the corresponding cross-section ratios, is
of the order of 6\% over the full range of $x$.         The uncertainty on the parametrisation
of $R$  affects the depolarisation factor $D$ [Eq.~(2)] by 4--5\%. The neglect of the $A_2$ term mainly
affects the highest $x$ interval where its contribution is estimated to be $\le 0.005$. The error on the
radiative corrections to the asymmetry is estimated by varying the input parametrisation of
$A_1^d(x)$  within the statistical error of the present data. 
The effect of event migration to  neighbouring $x$ bins, resulting from the
smearing of kinematic variables due to the finite resolution of the spectrometer and to the radiative effects,
was  evaluated by a Monte Carlo simulation and  found to be negligible. 
Potential false  experimental asymmetries were searched for by modifying the selection of
data sets used for the asymmetry calculation. 
The grouping of data into configurations with opposite target-polarisation
was  varied from large samples, covering at most two weeks of data taking, into about 100 small samples,
taken in time intervals of the order of 16 hours.
A statistical test was  performed on the distributions of the asymmetries extracted from these small samples.
In every  interval of $x$ they
were found to be normally distributed, with a standard deviation $\sigma$ compatible with the one derived from the
statistical errors ($\sigma_{stat}$).
Time-dependent effects which would lead to a broadening of these distributions were thus not observed.
Since the spread  of the observed $\sigma$'s is about 0.05, we take $1.1~\sigma_{stat}$ as upper limit 
for $\sigma$ and obtain  
for each $x$ bin a conservative upper bound of the 
systematic error arising from time-dependent effects
\begin{equation}
 \sigma_{syst} <  0.5~\sigma_{stat}.
\end{equation} 
 Asymmetries  for configurations where spin effects cancel out  were
calculated to check the cancellation of fluxes and acceptances. They were found compatible
with zero within their statistical errors. 
 The comparison of asymmetries obtained from different parts
of the spectrometer did not show any systematic effect.
Asymmetries obtained
with different settings of the DNP microwave frequency were compared  
in order to test possible effects related to the orientation of the target field.
No sizeable effect
was observed.

The values of $g_1^d(x,Q^2)$ quoted in the last column of Table~1 were     
obtained from Eq.~(5), with the $F_2^d$ parametrisation of
Ref.~\cite{smc} and the parametrisation of $R$ already used in the calculation of the depolarisation factor.
The systematic errors on $g_1^d$ contain an additional  contribution
due to the uncertainty on the parametrisation of $F_2^d$. The error due to the uncertainty on $R$ is 
reduced by a partial cancellation between the $R$ dependence of the depolarisation factor [Eq.~(2)]
and the factor $(1+R)$ in Eq.~(5).
 Our values of $g_1^d$ are shown in Fig.~\ref{fig:g1compa} in comparison with the SMC results
\cite{smc} which cover the same $Q^2$ range and were evolved to the same $Q^2$ values. 
Their improved
accuracy  provides a better evaluation of $g_1$ at low $x$:
integrating the values of $g_1^d(x)$ shown in   Fig.~\ref{fig:g1compa} over the range $0.004 < x < 0.03$,
we obtain $(-0.3 \pm 1.0) \cdot 10^{-3}$ and $(-5.3 \pm 2.3) \cdot 10^{-3}$ for COMPASS and SMC data,
respectively.
For $x <  0.03$ the COMPASS results are consistent with zero 
and do not show the tendency of the SMC data of negative $g_1^d$ values.

 In combination 
with the accurate SLAC and HERMES data at larger $x$, our new  results will improve the extrapolation
of $g_1^d$ towards $x = 0$. However,  taken alone, they do not provide 
a more accurate evaluation of the first moment $\Gamma_1^d$
because of the relatively large errors  at high $x$ resulting from the late implementation of the calorimetric trigger 
in the present data.
These errors will be reduced for the 2004 data where the calorimetric trigger was used during the full data-taking
period.    

In conclusion, a new evaluation of the longitudinal spin asymmetry 
and the spin structure function of the deuteron in the
DIS region ($Q^2 > 1~\GeV^2$) was  performed by the COMPASS experiment at CERN.
The data cover nearly the same range of $x$ as the former SMC experiment, $0.004 \le x \le 0.7$.   
The results are in  agreement with previous experiments over the full range of $x$ and
significantly improve the statistical accuracy in the region $x < 0.03$.

\section*{Acknowledgements}
We gratefully acknowledge the support of the CERN management and staff and  the skill and effort of
the technicians of our collaborating institutes.
Special thanks are due to V.~Anosov, J.M.~Demolis and V.~Pesaro for their technical support during
the installation and the running of this experiment.
 This work was made possible by the financial support of our funding agencies.

\end{document}